\documentclass[12pt]{iopart}

\usepackage{graphicx}
\usepackage{bm}
\usepackage{epstopdf}
\begin{document}

\title[]{Chimera states in nonlocally coupled phase oscillators with biharmonic interaction}
\author{Hongyan Cheng, Qionglin Dai, Nianping Wu, Yuee Feng, Haihong Li, and Junzhong Yang$^{*}$}
\address{School of Science, Beijing
University of Posts and Telecommunications, Beijing, 100876,
People's Republic of China}
\ead{jzyang@bupt.edu.cn}

\begin{abstract}
Chimera states, which consist of coexisting domains of coherent
and incoherent parts, have been observed in a variety of systems.
Most of previous works on chimera states have taken into account
specific form of interaction between oscillators, for example
sinusoidal coupling or diffusive coupling. Here, we investigate
chimera dynamics in nonlocally coupled phase oscillators with
biharmonic interaction. We find novel chimera states with features
such as that oscillators in the same coherent cluster may split
into two groups with a phase difference between them at around
$\pi/2$ and that oscillators in adjacent coherent clusters may
have a phase difference close to $\pi/2$. The different impacts of
the coupling ranges in the first and the second harmonic
interactions on chimera dynamics are investigated based on the
synchronous dynamics in globally coupled phase oscillators. Our
study suggests a new direction in the field of chimera dynamics.
\end{abstract}


 \maketitle

\section{Introduction}Chimera states refer to a type of spatiotemporal
pattern in which identical oscillators spontaneously split into
coexisting and spatially separated domains with dramatically
different behaviors, i.e., coherent and incoherent oscillations.
Chimera states were first numerically found in a ring of
nonlocally coupled Ginzburg-Landau oscillators \cite{kura}. Later,
Abrams and Strogatz presented theoretical results for the states
in a ring of phase oscillators coupled by a cosine kernel
\cite{abra}. Chimera states have been studied intensively over the
past years
\cite{abra08,lai09,mot10,nko13,set14,zak14,ban15,pazo14,cle16,lai15,she09,yel14,wolfrum}.
They have been found in the systems with different topologies,
such as square lattices \cite{mart10_a,gu13,li16}, torus
\cite{pana13}, and complex networks \cite{zhuy14}. The systems in
which chimera states are observed include time-discrete maps
\cite{omel11}, time-continuous chaotic models \cite{omel12},
neural systems \cite{omel13,hiza14,ber16}, and so on. Recently,
chimera states have been realized experimentally in optical
\cite{hage12,vik14}, chemical \cite{tins12,sch14}, mechanical and
electronic systems \cite{mart13,kap14,oml15}.

Generally, the system of nonlocally coupled identical oscillators
on a ring with length $L$ can be described as
\begin{eqnarray}\label{eq:1}
\dot{\mathbf{u}}(x)=\mathbf{f}[\mathbf{u}(x)]+\epsilon\int_{0}^{L}G(x-x')\mathbf{H}[\mathbf{u}(x)-\mathbf{u}(x')]dx'.
\end{eqnarray}
The kernel functions widely used are exponential function
$G(x)=e^{-\kappa |x|}$ and step function $G(x-x')=1$ if
$|x-x'|\leq\sigma$ or $G(x-x')=0$ otherwise, where $\kappa$ and
$\sigma$ measure the coupling range. For phase oscillators,
$u(x)=\theta(x)$, the phase of oscillator at position $x$, is a
scalar variable and $f[u(x)]=\omega$ the natural frequency of
oscillators which can be set to be zero without the loss of
generality. In previous investigations on nonlocally coupled
oscillators, the interaction among oscillators always takes the
form of diffusion coupling which manifests itself as
$\sin[\theta(x)-\theta(x')+\alpha]$, the first order of $H$ in the
Fourier expansion, for phase oscillators. However, in more general
cases, $H$ should be approximated by a biharmonic coupling
function
$\epsilon_1\sin[\theta(x)-\theta(x')+\alpha_1]+\epsilon_2\sin2[\theta(x)-\theta(x')+\alpha_2]$
\cite{erm,izh,czol13,gold13}. In fact, sinusoidally coupled
systems are typically degenerate, so the inclusion of the second
harmonic interaction may lead to more generic bifurcation
behaviors.

In this paper, We investigate nonlocally coupled phase oscillators
with biharmonic interaction. Especially, we suppose that the first harmonic and the second harmonic interactions may have different coupling ranges. We show the existence of novel
chimera states with the features resulted from the biharmonic
interaction. We explore the impacts of the coupling ranges in the
first and the second harmonic interactions on chimera states. We
present the mechanism for these novel chimera states based on the
synchronization in globally coupled phase oscillators.

\begin{figure*}
\includegraphics[width=5in]{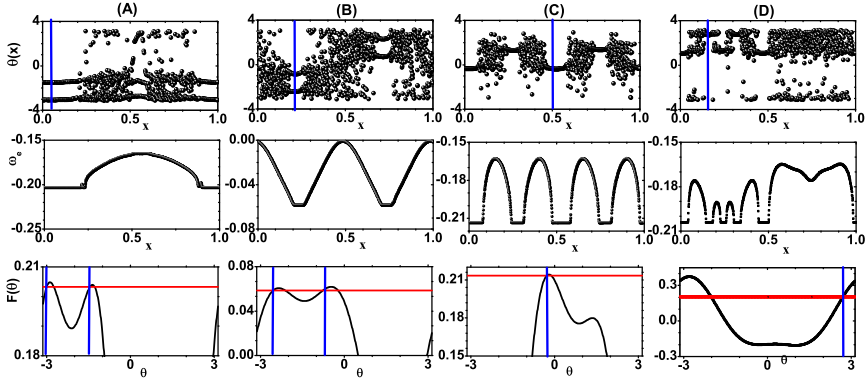}
\caption{\label{fig1}(color online) Examples of typical chimera
states in the model Eq.~(\ref{eq:2}). (A) The C1P2 chimera state
at $\kappa_1=4$ and $\kappa_2=4$ (B) The C2P2 chimera state at
$\kappa_1=2$ and $\kappa_2=4$. (C) The C4P1 chimera state at
$\kappa_1=4$ and $\kappa_2=0.5$. (D) The C5P1 chimera state at
$\kappa_1=4$ and $\kappa_2=1.4$. The top panels show the snapshots
$\theta(x)$ and the middle panels show the corresponding effective
frequencies $\omega_e(x)$. In the bottom panels, the black curve
plots the function $F(\theta)$ where the order parameters
$Z_{1,2}$ are acquired from the oscillator at the position denoted
by the blue line in the top panel and the red line plots $-\Omega$
the effective frequency in the coherent clusters. The blue lines
in the bottom panel denote the stable solutions $\theta(x)$ to
Eq.~(\ref{eq:6}). $N=1000$.}
\end{figure*}

\section{Model}We consider a ring of identical phase oscillators with
nonlocal coupling with fixed length $L=1$. The equation of motion for the system is
described as
\begin{eqnarray}\label{eq:2}
\dot{\theta}(x)&=&-\sum_{m=1,2}\epsilon_m\int_{0}^{1}G_{m}(x-x')\nonumber\\
&\times&\sin m[\theta(x)-\theta(x')+\alpha_m]dx'
\end{eqnarray}
Here, the angles $\alpha_1$ and $\alpha_2$ are tunable parameters
that describe the phase shifts between oscillators at $x$ and $x'$
in the first and the second harmonic interactions, respectively.
$\epsilon_1$ and $\epsilon_2$ describe the corresponding coupling
strengthes. The kernel functions $G_{1,2}$, decaying exponentially
with the distance between the oscillators, take the form
\begin{eqnarray}\label{eq:3}
G_{1,2}(x)=e^{-\kappa_{1,2}|x|}.
\end{eqnarray}
$\kappa_1$ and $\kappa_2$ measure the coupling ranges in the first
and the second harmonic interactions. To be noted, we do not
normalize the kernel functions to have unit integral and,
actually, the normalization constants are absorbed into the
coupling strengthes.

We define two position-dependent complex order parameters
(generalized Daido order parameters \cite{dai94,dai96})
$Z_{m}(x)=R_{m}(x)e^{im\Theta_m(x)}$ with $m=1,2$ as
\begin{eqnarray}\label{eq:4}
R_{m}(x)e^{im\Theta_{m}(x)}=\int_{0}^{1}G_{m}(x-x')e^{im\theta(x')}dx'.
\end{eqnarray}
Then Eq.~(\ref{eq:2}) becomes
\begin{eqnarray}\label{eq:5}
\dot{\theta}(x)&=&-\sum_{m=1}^{2}\epsilon_mR_m(x)\sin
m[\theta(x)-\Theta_m(x)+\alpha_m]
\end{eqnarray}
Now, the mutual entrainment among oscillators depends on both
$Z_1$ and $Z_2$.


Before going further, there are several remarks on the model.
Firstly, the model at $\kappa_1=0$ and $\kappa_2=0$ is reduced to
a globally coupled one. It has been investigated recently for
nonidentical phase oscillators with $\alpha_{1,2}=0$, where the
presence of the second harmonic interaction term leads to an
infinite number of coherent states \cite{koma13,li14}. On the
other hand, the model with the limit $\kappa_1\rightarrow\infty$
and $\kappa_2\rightarrow\infty$ represents the locally coupled
one. Secondly, both the model with $\epsilon_1\neq0$ and
$\epsilon_2=0$ and the model with $\epsilon_1=0$ and
$\epsilon_2\neq0$ support chimera states. In these two cases,
chimera states coexist with the synchronous state and realizing a
chimera state requires special initial conditions. There are an
infinite number of chimera states in the case of $\epsilon_1=0$
and $\epsilon_2\neq0$ due to the invariance of the model under the
transformation $\theta(x)\rightarrow\theta(x)+\pi$ for any $x$.
Thirdly, a similar version of the model Eq.~(\ref{eq:2}) has been
studied by Suda and Okuda where the kernel function is a step
function \cite{suda15}. They considered the situation with
$\epsilon_2\ll\epsilon_1$ and found two critical $\epsilon_2$. At
the first critical $\epsilon_2$, the synchronous state becomes
unstable and, then, chimera states may live forever. Beyond the
second critical $\epsilon_2$, chimera states are unstable. In this
paper, we will show that chimera states can exist for larger
$\epsilon_2$ and these states display novel features induced by
the second harmonic interaction. Ashwin and his collaborators
studied the model with biharmonic interaction in small systems and
proposed the concept of weak chimeras \cite{ash15,bick16,bick16a}.

\section{Results}We numerically simulate the model using a fourth-order
Runge-Kutta method with time step $\delta t =0.05$. The results
are examined by shorter time step such as $\delta t=0.01$. The
ring is discretized into $N$ oscillators with $N=256$ or $N=1000$.
Throughout the paper, $\alpha_1=1.45$ and $\alpha_2=1.45$. Without
loss of generality, we set $\epsilon_1=1$. We first let
$\epsilon_2=4$ to show the phenomenology of chimera states unique
to the model with the biharmonic interaction and to investigate
their dependence on the coupling ranges $\kappa_1$ and $\kappa_2$.
Then we present the dependence of the critical $\epsilon_2$ on
$\kappa_2$ above which these chimera states exist.

\begin{figure*}
\includegraphics[width=5in]{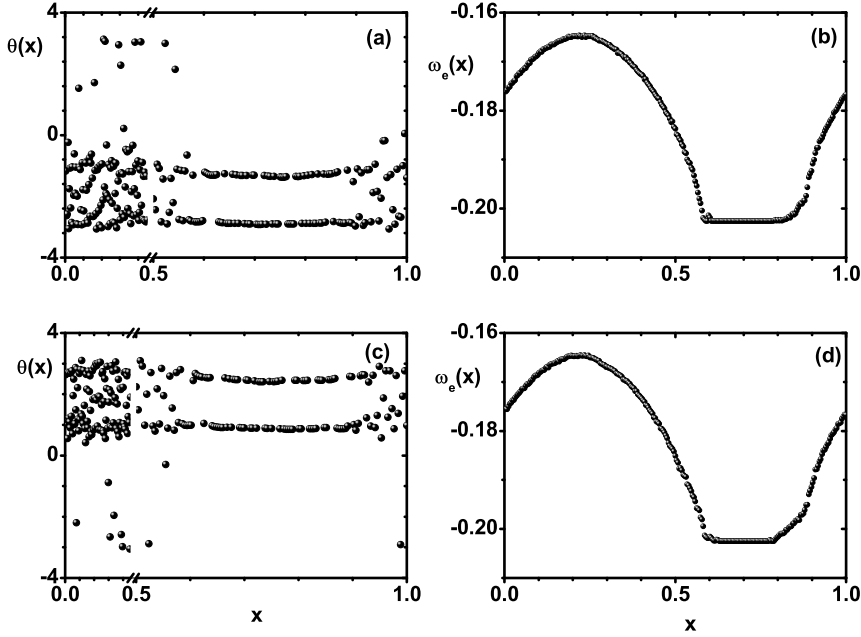}
\caption{\label{fig_2}(color online) Two C1P2 chimera states at
$\kappa_1=4$ and $\kappa_2=4$ generated from different random
initial conditions. $N=256$. (a) and (c) The snapshots
$\theta(x)$; (b) and (d) The corresponding effective frequencies
$\omega_e(x)$. Though both $\theta(x)$ and the profiles of
$\omega_e(x)$ for these two chimera states look the same, the
partitions of oscillators into two groups in the coherent cluster
are different from each other.}
\end{figure*}

\begin{figure*}
\includegraphics[width=6in]{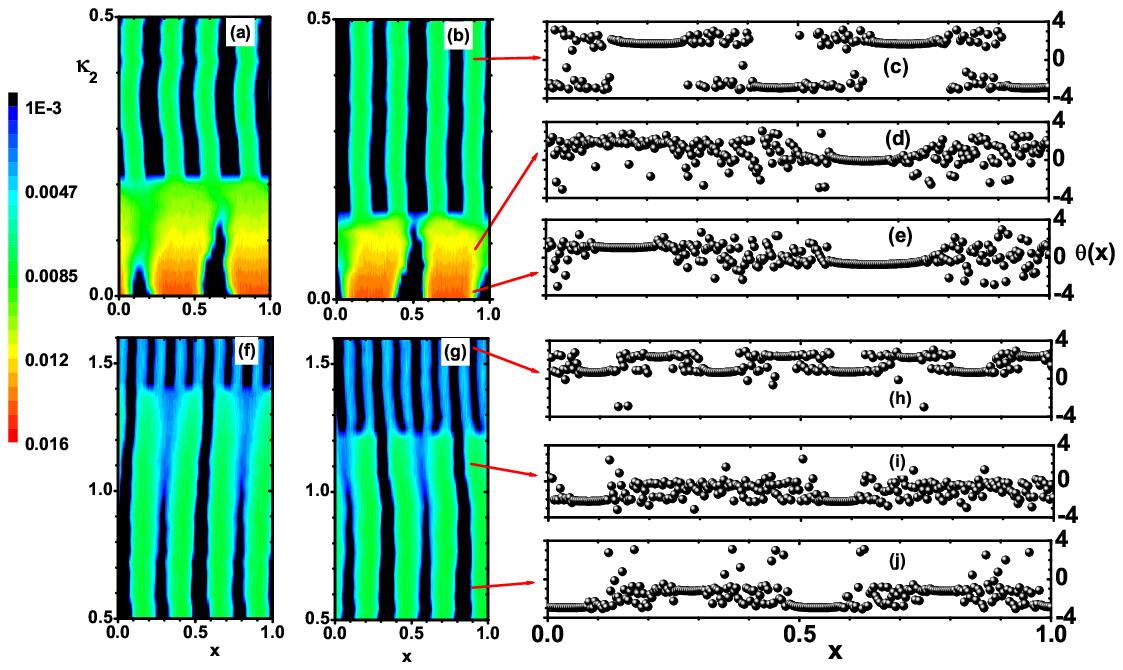}
\caption{\label{fig2}(color online) The forward and backward
transition diagrams between the C2P1 and C4P1 chimeras are
presented in (a) and (b) by monitoring $\sigma(x)$, respectively.
In these plots, coherent clusters are represented by black color
where $\sigma(x)=0$. (c), (d) and (e) show the snapshots
$\theta(x)$ at $\kappa_2$ denoted by the red arrows. The forward
and backward transition diagrams between the C4P1 and C8P1
chimeras are presented in (f) and (g), respectively. (h), (i) and
(j) show the snapshots $\theta(x)$ at $\kappa_2$ denoted by the
red arrows. $\kappa_1=4$ and $N=256$.}
\end{figure*}

\subsection{Typical chimera states}We focus on the parameter regime
$(\kappa_1,\kappa_2)\in[0,4]\times[0,4]$. In most part of the
regime, chimera states are stable provided that $\kappa_1$ is not
too close to zero, and can be realized with random initial
conditions where $\theta(x)$ is randomly drawn from $[0,2\pi]$.
There exist many types of chimera states. Some typical chimera
states are presented in Fig.~\ref{fig1} and discussed as
follows.

1) \textit{C1P2 chimera states} A chimera state at $\kappa_1=4$
and $\kappa_2=4$ is presented in Fig.~\ref{fig1}(A) where the
snapshot $\theta(x)$ and the effective frequencies $\omega_{e}(x)$
of oscillators are plotted in the top and the middle panels,
respectively. Here, the effective frequency $\omega_{e}(x)$ is
defined as $\omega_{e}(x)=\langle\frac{d\theta(x)}{dt}\rangle_{t}$
with $\langle\cdot\rangle_{t}$ the average over time. The graph
$\omega_e(x)$ shows clearly that oscillators split into two
clusters, one coherent cluster in which oscillators share the same
effective frequency and one incoherent cluster in which
oscillators have different effective frequencies. Different
$\omega_{e}(x)$ leads to scattered oscillators in the incoherent
cluster. However, different from the chimera states in previous
investigations where oscillators in the same coherent cluster are
nearly in phase, Fig.~\ref{fig1}(A) shows that the same
$\omega_e(x)$ does not lead oscillators to be nearly in phase. As
indicated by two nearly horizontal lines in the coherent cluster
in the graph $\theta(x)$, oscillators in the coherent cluster are
divided into two groups: oscillators in the same group are nearly
in phase and those in different groups have a phase difference
between them at around $\pi/2$. In the following, we denote as CnPm
a chimera state with $n$ coherent clusters and $m$ groups of
oscillators with different phases in a same coherent cluster is
Thus, the chimera state in Fig.~\ref{fig1}(A) is a C1P2 state.

The P2 phenomenon in a coherent cluster results from the
biharmonic interaction. To be clear, we consider oscillators in
the coherent cluster. For these oscillators, their phases are
locked to the mean fields and the phase differences
$\theta(x)-\Theta_{1,2}(x)$ between theirs and the mean fields can
be obtained by looking for the stable equilibria in
Eq.~(\ref{eq:5}). By using $\dot{\theta}(x)=\Omega$ for these
oscillators, Eq.~(\ref{eq:5}) yields
\begin{eqnarray}\label{eq:6}
-\Omega=F[\theta^*(x)]
\end{eqnarray}
with $F[\theta(x)]=\sum_{m=1,2}\epsilon_mR_m(x)\sin
m[\theta(x)-\Theta_m(x)+\alpha_m]$. The phase differences
$\theta^*(x)-\Theta_{1,2}(x)$ between these oscillators and the
mean fields $Z_{1,2}$ can be obtained by solving Eq.~(\ref{eq:6})
[To be noted, both $\theta^*(x)$ and $\Theta_{1,2}(x)$ are
time-dependent]. For $\epsilon_2=0$, Eq.~(\ref{eq:6}) has two
solutions and only one of them is stable. Therefore, there is only
one phase which can be taken by an oscillator in the coherent
cluster at any time, which is the reason behind that oscillators
in the coherent cluster stay nearly in phase and is what we have
observed in chimera states previously investigated. However, for
the biharmonic interaction with $\epsilon_2\neq0$, it is likely
that there are four different $\theta^*(x)-\Theta_{1,2}(x)$
satisfying Eq.~(\ref{eq:6}) and two of them are stable equilibria
to Eq.~(\ref{eq:5}), which means that there are two different
phases to be taken by an oscillator in the coherent cluster.
Consequently, oscillators in the coherent cluster spilt into two
groups and oscillators in different groups fall onto different
equilibria. The possible bistability in Eq.~(\ref{eq:5}) provides
an explanation on the P2 phenomenon. That the bistability does
exist in the chimera state in Fig.~\ref{fig1}(A) is examined in
the bottom panel where we consider an oscillator whose location is
indicated by the blue line in the top panel. $F(\theta)$ against
$\theta$ for the oscillator is presented with $Z_{1,2}$ and
$\Omega$ acquired in the simulation. As shown in the plot, there
exist four solutions and two of them, denoted by blue lines, are
stable. Comparing with the top panel, we find that two stable
$\theta^*(x)$ are the same as the phases taken by the oscillator.

In addition, which group an oscillator in the coherent cluster
belongs to depends on initial conditions. Therefore, the
oscillators in the coherent cluster randomly distribute themselves
into these two groups. For different initial conditions, the
2-group partitions are different. For example, we present in
Fig.~\ref{fig_2} two C1P2 states at $\kappa_1=4$ and $\kappa_2=4$
with different initial conditions. The snapshots $\theta(x)$ show
different partitions though their graphs $\omega_e(x)$ look the
same. If different partitions of oscillators in the coherent
cluster refer to different chimera states, there are a large
number of C1P2 states.

\begin{figure*}
\includegraphics[width=4in]{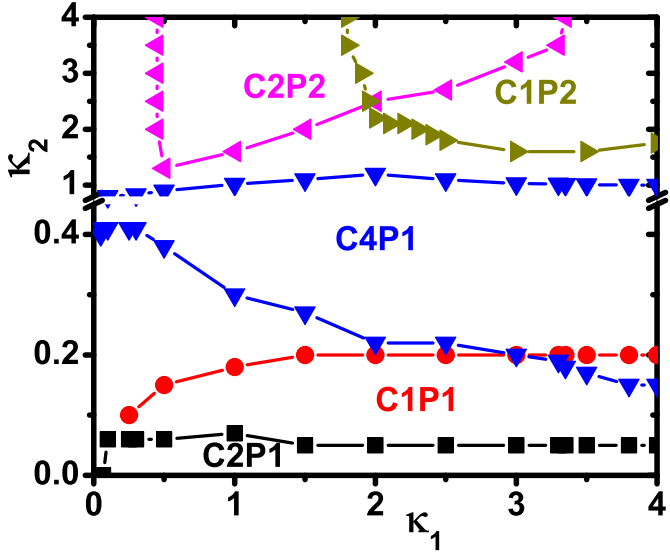}
\caption{\label{fig3}(color online) Stability regimes for several
typical chimera states in the plane of $\kappa_1$ and $\kappa_2$.
The C2P2 and C1P2 states are stable above the magenta and dark
yellow curves, respectively. The stable C4P1 is enclosed by the
blue curves. The C1P1 state is stable between the red and black
curves while the C2P1 state is stable below the black curve.
$N=256$. }
\end{figure*}

2) \textit{C2P2 chimera states}. As shown in Fig.~\ref{fig1}(B),
the state at $\kappa_1=2$ and $\kappa_2=4$ consists of two
coherent clusters separated by incoherent ones. In each coherent
cluster, oscillators split into two groups with a phase difference
between them at around $\pi/2$. Moreover, the oscillators in
different coherent clusters have a phase difference at around
$\pi$. The two coherent clusters always have the same size. Except
for the two groups of oscillators with different phases in each
coherent cluster, a C2P2 state looks similar to the clustered
chimera states in delay-coupled phase oscillators without the
biharminic interaction \cite{seth08}.

3) \textit{Multi-cluster chimera states with translation
symmetry}. This type of chimera states always manifest themselves
as C2nP1 or CnP1 states and have a spatial period $1/n$. A typical
C2nP1 state consists of $2n$ coherent clusters interspersed by
$2n$ incoherent ones and adjacent coherent clusters always have
different sizes. In each coherent cluster, oscillators are nearly
in phase. Different from previously studied multi-cluster chimera
states \cite{omel14,mais14}, there is a phase difference at around
$\pi/2$ between the adjacent coherent clusters instead of $\pi$.
In Fig.~\ref{fig1}(C), we show a C2nP1 state with $n=2$ at
$\kappa_1=4$ and $\kappa_2=0.5$. As indicated by the bottom panel,
there is only one stable equilibrium to Eq.~(\ref{eq:5}) for
oscillators in coherent clusters. On the other hand, a typical
CnP1 chimera state with spatial period $1/n$ consists of $n$
coherent clusters interspersed by $n$ incoherent ones. In CnP1
chimera states, adjacent coherent clusters always have the same
size and are nearly in phase between them. As shown later, CnP1 chimera
states with spatial period $1/n$ have close relation with C2nP1
states.

4) \textit{Multi-cluster chimera states with reflection symmetry}.
This types of chimera states consist of C2n+1P1 and C2n+1P2
states. In the states, coherent clusters always have different
sizes and adjacent coherent clusters have a phase difference at
$\pi/2$. The states own one largest incoherent cluster whose size
may be larger than 1/2. The states are symmetrical about the
center of the largest incoherent cluster. Figure~\ref{fig1}(D)
shows a C2n+1P1 chimera state with n=2 at $\kappa_1=4$ and
$\kappa_2=1.4$.

\begin{figure*}
\includegraphics[width=5in]{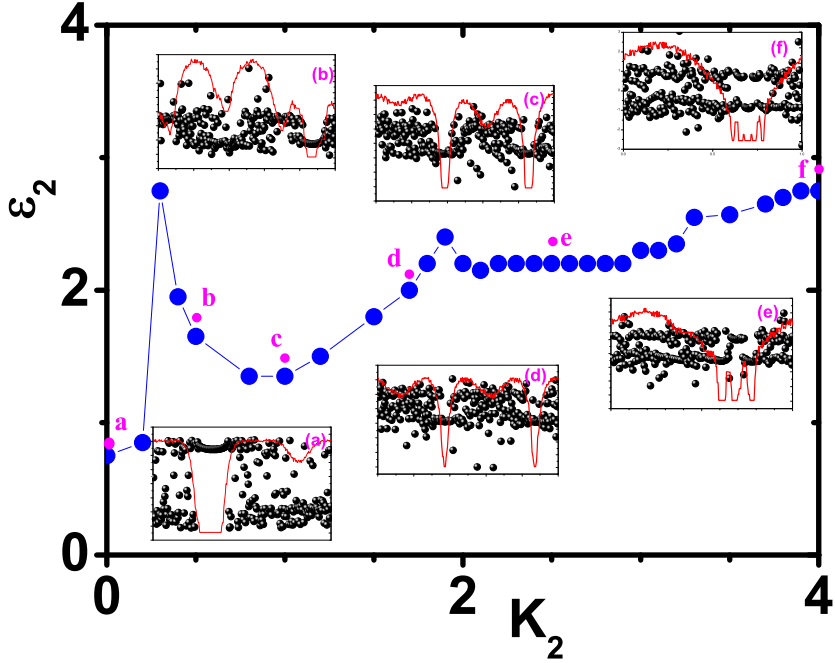}
\caption{\label{fig4}(color online)The critical $\epsilon_2$ is
plotted against $\kappa_2$. Above the curve, the chimera states
with the characteristics resulted from the biharmonic interaction
are born. Insets (a)-(f) show the snapshots $\theta(x)$ (black
dots) and the effective frequencies $\omega_e(x)$ (red curves) at
parameter sets denoted by magenta dots. (a) $\kappa_2=0$ and
$\epsilon_2=0.8$; (b) $\kappa_2=0.5$ and $\epsilon_2=1.73$; (c)
$\kappa_2=1$ and $\epsilon_2=1.5$; (d) $\kappa_2=1.7$ and
$\epsilon_2=2.2$; (e) $\kappa_2=2.5$ and $\epsilon_2=2.4$; (f)
$\kappa_2=4$ and $\epsilon_2=2.9$. $\kappa_1=4$ and $N=256$.}
\end{figure*}

\subsection{Stability regimes of typical chimera states}
We first explore the stability regimes of the typical chimera
states in the plane of $\kappa_1$ and $\kappa_2$. To do it, we
consider two transition diagrams with $\kappa_2$, forward and
backward continuations. The coupling range $\kappa_2$ is
successively increased/decreased by a $\delta\kappa_2$ in the
forward/backward continuation and the initial conditions for one
$\kappa_2$ are the final state of the previous one. During the
forward/backward continuation, the model evolves about $10^3$ time
units for each $\kappa_2$ to ensure that the steady state is
reached. When the transition such as the one between C2P1 states
and C4P1 states is taken into consideration, we first build a C2P1
state or a C4P1 state with random initial conditions and, then,
start the fowrward/backward continuation using the developed
C2P1/C4P1 state as the initial conditions. We monitor the quantity
$\sigma(x)= \langle[\dot\theta(x)-\omega_e(x)]^2\rangle_t$.
Generally, $\sigma(x)$ is zero in the coherent clusters for
stationary chimera states \cite{zhu12}, which can be used as a
criterion distinguishing chimera states with different numbers of
coherent clusters. Figures~\ref{fig2}(a) and (b) show the two
transition diagrams between C2P1 and C4P1 states at $\kappa_1=4$.
Clearly, the transition between C2P1 and C4P1 states is not
continuous. When $\kappa_2$ increases from $0$ to $0.5$, the C2P1
state first evolves into a C1P1 state in which the smaller
coherent cluster in the C2P1 one is lost, and then into a state in
which all oscillators become desynchronized. The C4P1 state shows
up suddenly when $\kappa_2$ is beyond $\kappa_2=2.1$. As an
illustration of chimera dynamics, the snapshots $\theta(x)$ at
different $\kappa_2$ are displayed in Figs.~\ref{fig2}(c)-(e). On
the other hand, the backward continuation shows that the C4P1
state persists till a lower $\kappa_2$ at around $1.5$ and, then,
the C1P1 state pops up abruptly which evolves into a C2P1 state
gradually. Figures~\ref{fig2}(f) and (g) show the transition
diagrams between C4P1 and C8P1 states at $\kappa_1=4$. As shown in
the figures, the transition between C4P1 and C8P1 is also
discontinuous and there exist C2P1 chimera states with spatial
period $1/2$ between them. The snapshots of these different
chimera states are presented in Figs.~\ref{fig2}(h)-(j).

Basing on the transition diagrams at different $\kappa_1$, we have
the stability regimes in the plane of $\kappa_1$ and $\kappa_2$
for typical chimera states in Fig.~\ref{fig3}. The C2P1 state is
always stable at $\kappa_2=0$ provided that $\kappa_1>0.06$. The
regime for the C2P1 state is confined to a narrow regime around
$\kappa_2=0$ with a threshold at around $\kappa_2=0.05$. The
stability regime of the C4P1 state, which is enclosed by the blue
curves in Fig.~\ref{fig3}, is separated from that of the C2P1
state by the chimera state C1P1. Interestingly, we find that the
stabilities of the C2nP1 states with the spatial period $1/n$
strongly depend on $\kappa_2$ but are not sensitive to $\kappa_1$.
On the other hand, the stabilities of the C1P2 and C2P2 chimera
states are sensitive to $\kappa_1$. As shown, the C1P2 chimera
state requires large $\kappa_1$ while the C2P2 one prefers
intermediate $\kappa_1$. Even if only a few types of chimera
states are taken into considerations, the overlap amongst the
stability regimes is apparent in Fig.~\ref{fig3}, which
suggests that the coexistence among different types of chimera
states is prevailing in the model Eq.~(\ref{eq:2}).

As mentioned in the model section, to realize the chimera states
with the features resulting from the biharmonic interaction, the
strength of the second harmonic interaction has to be strong
enough. Then, we consider chimera dynamics in the plane of
$\kappa_2$ and $\epsilon_2$. Instead of the stability regimes of
different chimera states, we concern with the onset of the chimera
states observed above with the change of $\epsilon_2$. For a given
$\kappa_2$, we simulate the model Eq.~(\ref{eq:2}) with random
initial conditions at different $\epsilon_2$. For each
$\epsilon_2$, we perform simulations tens of times. If no chimera
state can be developed,  $\epsilon_2$ is below the critical value.
Following this way, we locate the critical $\epsilon_2$ for
different $\kappa_2$ and the curve of the critical $\epsilon_2$
against $\kappa_2$ is presented in Fig.~\ref{fig4}. The critical
$\epsilon_2$ does not change monotonically with $\kappa_2$. At
around $\kappa_2=0.5$, the critical $\epsilon_2$ jumps to a high
value. In the range of $\kappa_2\in(0.5,2)$, the curve behaves
like a parabolic one. Furthermore, the critical $\epsilon_2$
increases with $\kappa_2$ when $\kappa_2>2$. Interestingly, in
these extensive simulations, we find that, with the increase of
$\epsilon_2$ from the critical value, $CnP1$ chimera states with
spatial period $1/n$ are born firstly for $\kappa_2<2$ while
multi-cluster chimera states with reflection symmetry are born
firstly for $\kappa_2>2$. Some examples of chimera states at the
parameters just above the critical curves are presented in the
insets in Fig.~\ref{fig4}.

\begin{figure*}
\includegraphics[width=6in]{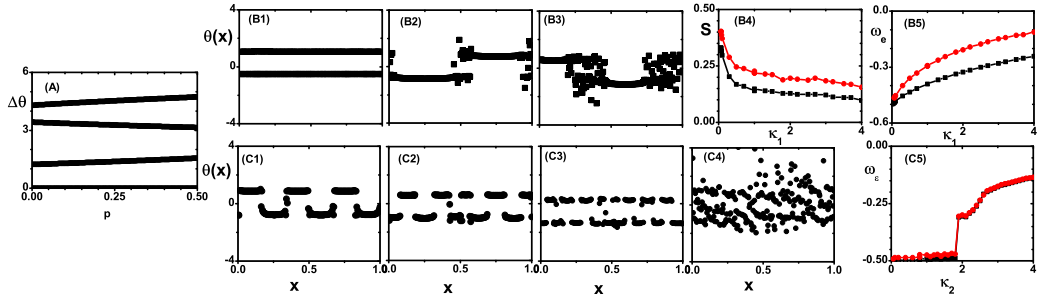}
\caption{\label{fig5}(color online) (A) The solutions
$\Delta\theta$ to Eq.~(\ref{eq:7}) are plotted against $p$ where
$\Delta\theta=0$ and $\Delta\theta=\pi$ are unstable. (B) The
effects of $\kappa_1$ on chimera dynamics at $\kappa_2=0$.
(B1)-(B3) The snapshots $\theta(x)$ at $\kappa_1=0.04$,
$\kappa_1=0.07$ and $\kappa_1=0.5$, respectively. (B4) The sizes
of two coherent clusters in the C2P1 states are plotted against
$\kappa_1$. (B5) The maximum (in red) and the minimum (in black)
in the graph $|\omega_e(x)|$ are plotted against $\kappa_1$. (C)
The effects of $\kappa_2$ on chimera dynamics at $\kappa_1=0$.
(C1)-(C4) The snapshots $\theta(x)$ at $\kappa_2=0.1$,
$\kappa_2=0.2$, $\kappa_2=0.8$, and $\kappa_2=2$, respectively.
(C5) The maximum (in red) and the minimum (in black) in the graph
$|\omega_e(x)|$ are plotted against $\kappa_2$. N=256.}
\end{figure*}


\subsection{The link between chimera states and synchronization in globally coupled model}
To better understand the above chimera dynamics, it is helpful to
investigate the synchronous dynamics in globally coupled
oscillators ($\kappa_1=0$ and $\kappa_2=0$). In such a synchronous
state, oscillators will split into two groups and, in each of
them, oscillators hold a same phase. Denoting $\theta_{i}$
($i=1,2$) as the phases held by oscillators in these two groups,
we have
\begin{eqnarray}\label{eq:7}
\Omega=&-&\epsilon_1R_1\sin(\theta_i-\Theta_1+\alpha_1)\nonumber\\
&-&\epsilon_2R_2\sin2(\theta_i-\Theta_2+\alpha_2), i=1, 2
\end{eqnarray}
with $\Omega=d\theta_1/dt=d\theta_2/dt$. Suppose that the fraction
of oscillators taking $\theta_1$ is $p$. Then substituting the
order parameters Eq.~(\ref{eq:3}) into Eq.~(\ref{eq:7}), we have
the relationship between $\Delta\theta=\theta_2-\theta_1$ and $p$,
which is shown in Fig.~\ref{fig5}(A) (A relevant theoretical work
has been done in the reference \cite{oro09}). At a given $p$,
there are four different $\Delta\theta$ satisfying
Eq.~(\ref{eq:7}) in which $\Delta\theta=0$ and
$\Delta\theta\simeq\pi$ are unstable. The other two $\Delta\theta$
are stable: $\Delta\theta\simeq\pi/2$ and
$\Delta\theta\simeq-\pi/2$ (or $\Delta\theta\simeq3\pi/2$).
Interestingly, the values of these two stable $\Delta\theta$ are
close to both the phase difference between the two groups in
coherent clusters for C1P2 and C2P2 chimeras and the phase
difference between adjacent coherent clusters for C2nP1 chimeras
with spatial period $1/n$.

To elucidate the relation between the synchronous dynamics in the
case of global coupling and the chimera dynamics in the case of
nonlocal coupling and to explore the impacts of $\kappa_1$ and
$\kappa_2$ on chimera dynamics, we investigate the model
Eq.~(\ref{eq:2}) by increasing either $\kappa_1$ or $\kappa_2$
from zero. Figures~\ref{fig5}(B1) and (B2) show the snapshots
$\theta(x)$ for two nonzero $\kappa_1$ at $\kappa_2=0$. At small
$\kappa_1$, the synchronous dynamics persists. Beyond a critical
$\kappa_1$ which depends on $p$, the synchronous state becomes
unstable and oscillators self-organize themselves into a C2P1
chimera state during which the phase difference between
oscillators in synchronization keeps almost unchanged. Further
increasing $\kappa_1$ does not alter the nature of the C2P1 state
[see Fig.~\ref{fig5}(B3)]. Figure~\ref{fig5}(B4) shows the sizes
of coherent clusters $S$ against $\kappa_1$. The sizes of both
coherent clusters decrease with $\kappa_1$ monotonically. On the
other hand, the size difference between two coherent clusters
seems to be independent of $\kappa_1$. We also monitor the maximum
and the minimum in the graph $|\omega_e(x)|$. The results
presented in Fig.~\ref{fig5}(B5) show that the decrease of the
fraction of coherent oscillators with $\kappa_1$ is accompanied by
the fall of $|\omega_e(x)|$. In contrast, $\kappa_2$ displays
quite different impacts on the model dynamics.
Figures~\ref{fig5}(C1)-(C3) suggest that increasing $\kappa_2$
leads to more coherent clusters. At sufficient large $\kappa_2$,
the states with many coherent clusters are replaced by irregular
dynamics [see Fig.~\ref{fig5}(C4)]. Figure~\ref{fig5}(C5) shows
that the maximum and the minimum in the graph $|\omega_e(x)|$
changes prominently only for irregular dynamics.

In short, the results in Fig.~\ref{fig5} suggest that the
synchronous dynamics in globally coupled phase oscillators is
responsible for the $\pi/2$ phenomenon for both the phase
difference between two groups of coherent oscillators in C1P2 and
C2P2 chimera states and the phase difference between adjacent
coherent clusters in chimera states such as C2nP1 ones.
Figure~\ref{fig5} also suggests that it is the coupling range
$\kappa_2$ but not the coupling range $\kappa_1$ to determine the
number of coherent clusters in CnP1 chimera states, which is in
agreement with Fig.~\ref{fig3}. In addition, Fig.~\ref{fig5} shows
that synchronous dynamics in the model Eq.~(\ref{eq:2}) are always
unstable provided that $\kappa_1$ is not close to zero. Without
the competition with the synchronous dynamics, chimera states in
the presence of the biharmonic interaction can always be realized
with arbitrary initial conditions, which is quite different from
the chimera states in the absence of the second harmonic
interaction.

\section{Conclusion}In conclusion, we have investigated the
nonlocally coupled phase oscillators with the biharmonic
interaction. We found chimera states with peculiar characteristics
resulting from the interplay between the first and second harmonic
interactions. In C1P2 and C2P2 chimera states, oscillators in the
same coherent cluster spilt into two groups with the phase
difference between them at around $\pi/2$. In C2nP1 chimera states
with spatial period $1/n$, the phase difference between adjacent
coherent clusters is not $\pi$ but around $\pi/2$. We also found
the prevalence of multi-cluster chimera states. Multi-cluster
chimera states have been found in many systems with kernel
function taking the form of step function
\cite{omel13,omel14,hiza14}. In those works, reducing the coupling
range favors the appearance of more coherent clusters. However,
for kernel function taking the exponential one, only transient
multi-cluster chimera states are observed for phase oscillators
\cite{mais14}. Therefore, multi-cluster chimera states with more
than 2 coherent clusters are not trivial for nonlocally coupled
phase oscillators. Furthermore, we found that the coupling range
$\kappa_1$ has little influence on the number of coherent
clusters. In contrast, increasing the coupling range $\kappa_2$
seems to produce more coherent clusters when $\kappa_2$ is not
very large.

\section{Acknowledge}This work was supported by NSF of China under
Grants Nos. 11575036, 71301012, and 11505016.

\end{document}